\documentclass[aps,floats,amssymb,twocolumn,superscriptaddress,showpacs]{revtex4-1}
\usepackage{graphicx,subfig,amsmath,hyper ref}
\newcommand{\braket}[2]{\langle #1|#2\rangle}

\newcommand{\abs}[1]{|#1|}
\begin{document}
\begin{abstract}
We discuss $d$-wave topological (broken time reversal symmetry) pairing structures in unpolarized and polarized Jain states. We demonstrate pairing in the Jain spin singlet state by rewriting it in an explicit pairing form, in which we can recognize $d$-wave weak pairing of underlying quasiparticles - neutral fermions. We find and describe the root configuration of the Jain spin singlet state and its connection with neutral excitations of the Haldane-Rezayi state, and study the transition between these states via exact diagonalization. We find high overlaps with the Jain spin singlet state upon a departure from the hollow core model for which the Haldane-Rezayi state is the exact ground state. Due to a proven algebraic identity we were able to extend the analysis of topological $d$-wave pairing structures to
polarized Jain states and integer quantum Hall states, and discuss its consequences.
\end{abstract}


\title{Topological $d$-wave pairing structures in Jain states}
\author{N. Moran}
\affiliation{Laboratoire Pierre Aigrain, ENS and CNRS, 24 rue Lhomond, 75005
Paris, France}
\author{A. Sterdyniak}
\affiliation{Laboratoire Pierre Aigrain, ENS and CNRS, 24 rue Lhomond, 75005
Paris, France}
\author{I. Vidanovi\'c}
\affiliation{Scientific Computing Laboratory, Institute of Physics
Belgrade, University of Belgrade, Pregrevica 118, 11 080 Belgrade,
Serbia}
\author{N. Regnault}
\affiliation{Department of Physics, Princeton University, Princeton, NJ 08544}
\affiliation{Laboratoire Pierre Aigrain, ENS and CNRS, 24 rue Lhomond, 75005
Paris, France}
\author{M.V. Milovanovi\'c}
\affiliation{Scientific Computing Laboratory, Institute of Physics
Belgrade, University of Belgrade, Pregrevica 118, 11 080 Belgrade,
Serbia}
\maketitle

\section{Introduction}
Fractional quantum Hall (FQH) states are strongly correlated many-body states which in certain cases have an effective description in terms of weakly-interacting quasiparticles. An important example are Jain states \cite{book_jain} which are composed of weakly-interacting composite fermion quasiparticles, which themselves form underlying integer quantum Hall (IQH) states. In other important examples these underlying states of quasiparticles may be superconducting with broken time reversal symmetry, like in the famous Pfaffian (Moore-Read) state \cite{mr} with $p$-wave superconducting pairing of neutral fermion quasiparticles. The paired states in the FQHE are often discussed in connection with systems with extra degrees of freedom like spin. The first paired state proposed was the spin-singlet $d$-wave Haldane-Rezayi (HR) state \cite{hr}. It has served as inspiration and as a prototype for other paired states, despite  initial confusion about its compressibility. Initially it was believed to be an incompressible state - a spin-singlet state at filling factor 1/2. However in Ref. \onlinecite{rg} the HR state was identified as a critical (gapless) state of a $d$-wave superconductor with broken time reversal symmetry. In the same reference it was shown that the gapped phase that is on the weak pairing side of the transition for which the HR state is critical possesses some universal properties of the Jain spin singlet (JSS) state at half filling \cite{bwj}. Therefore the JSS state may represent a weakly-paired $d$-wave topological superconductor of neutral fermion quasiparticles and may be related to the gapless HR $d$-wave state. On the other hand, recent developments in the theory of the FQHE have demonstrated exceptional similarities between polarized Jain states and a
non-unitary series of states (connected with non-unitary conformal field theories (CFTs)) with gapless behavior \cite{ss,rgj,nr,bh,sb}.

In this paper we focus on $d$-wave topological pairing structures in unpolarized and polarized Jain states. First we discuss further the connection between the JSS state and topological $d$-wave superconductors, and the implied connection between HR and JSS states. Due to an algebraic identity we recover the exact pairing (structure) in the JSS wavefunction. The root configuration of the same state is also presented. These results improve
our understanding of the role of paired neutral fermions in the HR and JSS state, and the transition that is expected to occur between these states. In order to confirm its existence in the presence of specific interactions we study this transition by way of exact diagonalization. Due to the spin degree of freedom our studies are limited in the system sizes treated compared to studies without spin. In the systems we could treat we demonstrate high overlaps with the JSS state upon departing from the pure hollow core model for which the HR state is the exact ground state. Due to the proven identity we are able to show that the pairing structures also exist even in polarized Jain states, as a consequence of the underlying multicomponent nature of the FQH states. Furthermore we demonstrate a connection, based on the proven identity, between the IQH states with Chern number equal to two\cite{qi2011,wang2011,wang2012} and the $d$-wave superconducting states with broken time reversal symmetry. This connection is enabled by the extremely weak pairing in the $d$-wave superconductor. We will discuss the connection on the level of many-body wavefunctions; it was introduced previously on the level of Hamiltonians by Laughlin in Ref. \cite{lau}.

The paper is organized as follows: section \ref{sec:model_wfs} introduces the HR and JSS model wavefunctions and reviews their most relevant properties, section \ref{sec:pairing_structure} shows how to see hidden pairing structure in the JSS state, section \ref{sec:root_partitions} discusses the HR and JSS states in terms of their root partitions, section \ref{sec:num_calcs} presents results from numerical calculations, section \ref{sec:spin_polarised_case} extends the pairing structure arguments to the spin polarized case and finally section \ref{sec:conclusions} presents conclusions.

\section{Model wavefunctions}\label{sec:model_wfs}
To understand better the topological nature of Jain states and their relationship to the non-unitary states we will first discuss the JSS state and related HR state.
The JSS state at $\nu = \frac{1}{2}$ is defined as
\begin{equation}
\Psi_{JSS} = \mathcal{P}_{LLL}(\chi_2 \chi_{110} \chi_1)\label{first}
\end{equation}
in the usual Jain notation. $\mathcal{P}_{LLL}$ is the projector operator to the lowest Landau level (LLL). $\chi_2$ denotes the wavefunction of two filled Landau levels (LLs) of all particles. As shown in Ref. \cite{hh}, in a condensed form $\chi_2$ can be expressed as
\begin{eqnarray}
\chi_2 = {\cal A}\{ \prod_{i=1}^M z_i^* & \times & \prod_{i<j;i,j\le M} (z_i - z_j)\nonumber \\
& \times  & \prod_{k<l;M < k,l \le N} (z_k - z_l)\} \label{chi2},
 \end{eqnarray}
 where $N$, the total number of particles, is assumed even, and $M = N/2$. ${\cal A}$ denotes the antisymmetrization operator over the $N$ particles. Here and below we suppress the omnipresent Gaussian factors, characteristic of the disk
 geometry. The steep potential at the edge of the system for the disk geometry means that the wavefunction for two filled LLs has one less electron in the LLL than the second LL. Nevertheless we will neglect this edge uncertainty, because in this section we look for long-distance properties of wavefunctions, and use the expressions \ref{chi2} or \ref{det} (below).
$\chi_1$ denotes the wavefunction of a filled LLL of all particles.
\begin{equation}
\chi_1 = \prod_{i<j}^{N}(z_i - z_j) \label{chi1},
\end{equation}
and $\chi_{110}$ denotes the wavefunction with Jastrow-Laughlin factors only between particles with the same spin.
\begin{equation}
\chi_{110} = \prod_{i<j}^{\frac{N}{2}}(z_i^{\uparrow} - z_j^{\uparrow})\prod_{i<j}^{\frac{N}{2}}(z_i^{\downarrow} - z_j^{\downarrow})\label{chi110},
\end{equation}
where $z_i^{\uparrow}$ ($z_i^{\downarrow}$) are the positions of the particles with spin up (down). Where no spin index is given, the product is over all particles irrespective of spin.

The HR state \cite{hr} is a fermionic spin singlet state defined as
\begin{equation}
 \Psi_{HR} = \det\left(\frac{1}{(z_i^{\uparrow} - z_j^{\downarrow})^2}\right)\prod_{i < j}(z_i - z_j)^2.
\end{equation}
This state is the unique densest zero energy ground state of a hollow core two-body interaction Hamiltonian. Two-body interaction Hamiltonians can be expressed in terms of the Haldane pseudo-potential coefficients $V_m$ \cite{hpp} as
\begin{equation}
 H = \sum_{m \ge 0} \left( V_m \sum_{i < j}\mathcal{P}_{ij}^{(m)} \right),
\end{equation}
where $V_m$ is the pseudo-potential coefficient for relative angular momentum $m$ and $\mathcal{P}_{ij}^{(m)}$ projects a particle pair onto relative angular momentum $m$. The hollow core interaction corresponds to setting the $V_1$ coefficient to a finite value while the rest are set to zero. For the HR state the counting of zero modes with and without quasi-holes can be deduced from a generalized Pauli principle \cite{rr96,ar11}.

We will examine in detail the transition induced by changing $V_0$ (interaction pseudo-potential for particles with relative angular momentum zero) that is believed to represent the transition from HR to JSS state. We are especially interested in identifying the JSS and its universal properties on the weak pairing side of the transition. This will also entail better examination of the JSS along with its root configuration.

\section{Pairing structure}\label{sec:pairing_structure}
From the expression for the JSS state in (\ref{first}) we will illustrate the basic pairing structure that is hidden in the usual definition of Jain states.
We will prove an algebraic identity in this case that directly relates the JSS wavefunction and the long-distance form of the ground state of a $d$-wave topological superconductor in its weak pairing phase.

The projection to the LLL is made by replacing complex conjugate coordinates,
$z_i^*, i = 1, \ldots, N$ in the two LL filled wavefunction, $\chi_2$, with derivatives, $\partial/(\partial  z_i), i = 1, \ldots, N$. When attempting to construct this state numerically we found that changing the order of application of the projection operator to reduce the computational complexity is no longer applicable here as it is in the spin-less case  \cite{jk,jr}. For further details see appendix \ref{ap1}. We will use expression \ref{chi2} for $\chi_2$, derived in Ref. \cite{hh}, which assumes even numbers of particles, $N = 2 M$.
It is important to notice that in the equivalent but more common definition of $\chi_2$,

\begin{equation}
\chi_2 = \left|\begin{array}{ccccc}  1 & \cdots & 1 & \cdots & 1 \\
                                     z_1 & \cdots   & z_i & \cdots & z_N \\
                                    z_1^2 & \cdots & z_i^2& \cdots & z_N^2\\
                                    \vdots &  & \vdots & & \vdots\\
                                    z_1^{M-1} &\cdots& z_i^{M-1} & \cdots & z_N^{M-1}\\
                                    z_1^* & \cdots   & z_i^* & \cdots & z_N^* \\
                                    z_1^*z_1 & \cdots   & z_i^* z_i & \cdots & z_N^*z_N \\
                                    z_1^*z_1^2 & \cdots   & z_i^* z_i^2 & \cdots & z_N^*z_N^2 \\
                                    \vdots &  & \vdots & & \vdots\\
                                    z_1^* z_1^{M-1} &\cdots& z_i^*z_i^{M-1} & \cdots & z_N^* z_N^{M-1}\\
\end{array} \right|, \label{det}
\end{equation}
due to the asymmetry of the determinant, any exchange of two particles amounts only to a change of sign analogous to the wavefunction of a filled LLL, expression \ref{chi1}. If we use these expressions for two groups of particles as in the case of states with spin assignment, which particles are up or down becomes irrelevant (as far the correlations are concerned) as these expressions have equal correlations for up - up, down - down, and up - down correlators. It is important to notice that spin is not fixed in a given LL (in $\chi_2$ in the definition, expression \ref{chi2} or \ref{det}), and each LL may contain any distribution of ups and downs.  In the following we will extract (under derivatives due to the LLL projection) from each term in $\chi_2$ the correlator that is between the two definite groups with the same number of particles equal to $M$; the first group will be among particles to which we assign spin up and the second group will be among particles with spin down.
Therefore we have
\begin{eqnarray}
\Psi_{JSS} =&& {\cal A} [ \partial_{z_1} \cdots \partial_{z_M} \nonumber \\
&& \frac{\prod_{i<j;i,j\le M} (z_i - z_j)\prod_{k<l;M < k,l \le N} (z_k - z_l)}{\prod_{p,q} (z_{p \uparrow} - z_{q \downarrow})} ] \nonumber \\
&& ( \prod_{p,q} (z_{p \uparrow} - z_{q \downarrow})] \chi_{110} \chi_1 ).
\label{jssquare}
\end{eqnarray}
Only if the division into two groups under ${\cal A}$ coincides with division between up and down particles can we use the Cauchy identity,
\begin{eqnarray}
\frac{\prod_{i<j} (z_{i \uparrow} - z_{j \uparrow})\prod_{l<m} (z_{l \downarrow} - z_{m \downarrow})}{\prod_{p,q} (z_{p \uparrow} - z_{q \downarrow})} = \det\left(\frac{1}{z_{p \uparrow} - z_{q \downarrow}}\right), \nonumber
\end{eqnarray}
where the resulting determinant has antisymmetry among same spin particles. This gives us a clue about what the expression under the square brackets in Eq.(\ref{jssquare}),
\begin{eqnarray}
&& {\cal A}[\partial_{z_1} \cdots \partial_{z_M} \nonumber \\
&&  \frac{\prod_{i<j;i,j\le M} (z_i - z_j)\prod_{k<l;M < k,l \le N} (z_k - z_l)}{\prod_{p,q} (z_{p \uparrow} - z_{q \downarrow})}],
\end{eqnarray}
should be.

The expression

(a) should not carry macroscopic flux (the filling factor is determined by $[ \prod_{p,q} (z_{p \uparrow} - z_{q \downarrow})] \chi_{110} \chi_1 = \chi_1^2$),

(b) should preserve the same total power $(N/2 = M)$ of derivatives,

(c) should be antisymmetric under exchange of same spin particles,

(d) and should be invariant under total (when all particles participate) exchange between opposite spin particles due to the factor $\prod_{p,q} (z_{p \uparrow} - z_{q \downarrow})$ that already encodes a definite symmetry of $\chi_2$ under the total exchange equal to the parity of $M^2$ i.e. $(-1)^{M^2}= (-1)^M$ between opposite spin particles,

(f) and should be invariant under translation (as  $\chi_2$ is).

This is achieved by the following pairing function,
\begin{equation}
\Psi_d = \det\left(\frac{z^*_{p \uparrow} - z^*_{q \downarrow}}{z_{p \uparrow} - z_{q \downarrow}}\right),\label{pair_function}
\end{equation}
to which the projection to the LLL has to be applied when considering the JSS state.

To see that the function is invariant under any total exchange between up and down particles we start with a general expression,
\begin{equation}
\Psi = \sum_{p\in \mathcal{S}_M} f_{1,p(2)} \cdots f_{2M-1,p(2M)} \textrm{sgn}(p),
\end{equation}
for a pairing function of $M$ pairs. $\mathcal{S}_M$ is the symmetric group over a set of $M$ elements and $\textrm{sgn}(p)$ is the signature of the permutation $p$. Each pair is invariant under the exchange of  its constituents i.e. $f_{i,j} = f_{j,i}$. Any total exchange between two kinds (even and odd) of particles is defined by a single permutation $s$ on $M$ numbers. The transformed wavefunction, ${\cal E}\Psi$, can be expressed as
\begin{eqnarray}
{\cal E}\Psi = && \sum_p f_{s^{-1}p(2),s(1)} \cdots f_{s^{-1}p(2M),s(2M-1)} \textrm{sgn}(p) \nonumber \\
        = && \sum_p f_{s(1),s^{-1}p(2)} \cdots f_{s(2M-1),s^{-1}p(2M)} \textrm{sgn}(p) \nonumber \\
        = && \sum_p f_{1,s^{-2}p(2)} \cdots f_{2M-1,s^{-2}p(2M)} \textrm{sgn}(p) \nonumber \\
        = && \sum_\sigma f_{1,\sigma(2)} \cdots f_{2M-1,\sigma(2M)} \textrm{sgn}(\sigma) = \Psi,
\end{eqnarray}
i.e. we proved that the pairing function is invariant under any total exchange ${\cal E}$ between (ups and downs) even and odd number particles.

Thus we have
\begin{eqnarray}
\Psi_{JSS} = && \det\left( \frac{\partial_{z_\uparrow} - \partial_{z_\downarrow}}{z_{\uparrow} - z_{\downarrow}}\right)[ \prod_{i,j} (z_{i \uparrow} - z_{j \downarrow})] \chi_{110} \chi_1 \nonumber \\
 = &&\det\left( \frac{\partial_{z_\uparrow} - \partial_{z_\downarrow}}{z_{\uparrow} - z_{\downarrow}}\right)
 \chi_1^2. \label{finaljss}
\end{eqnarray}
The existence and uniqueness of the pairing function that satisfies the listed conditions leads to the equality of expressions.
While we don't have a proof of the uniqueness of the pairing wavefunction, we checked the following identity
\begin{equation}
\chi_2 = \Psi_d   \prod_{i,j} (z_{i \uparrow} - z_{j \downarrow}), \label{identity}
\end{equation}
and thus Eq.(\ref{finaljss}), hold true up to $N \leq 8$.
Interestingly we came to an expression for $\chi_2$ that includes the division into two groups of particles, but as we emphasized previously this does not select any particular two groups in the definition of $\chi_2$ as long as we do not assign spin. But in the definition of the JSS wavefunction we do and it is then natural to decompose $\chi_2$ in a way that respects this spin assignment.

\section{Root partitions}\label{sec:root_partitions}

In the following we will describe another characteristic of the JSS, its root configuration. It has been established \cite{bh08} that many model FQH states can be written exactly as Jack polynomials or as the product of a Jack polynomial and some power of Vandermonde determinants. Jack polynomials are characterized by a dominant partition which reflects the vanishing properties of the state. A partition $\lambda$ can be represented as an occupation-number configuration $n(\lambda) = \{ n_m(\lambda), m = 0, 1, 2, ... \}$ of each of the LLL orbitals. A ``Squeezing rule'' connects configurations $n(\lambda) \to n(\mu)$. This is a two particle operation that moves a particle from orbital $m_1$ to $m_1^{\prime}$ and another from $m_2$ to $m_2^{\prime}$ with $m_1 < m_1^{\prime} <= m_2^{\prime} < m_2$ and $m_1 + m_2 = m_1^{\prime} + m_2^{\prime}$. A configuration $\lambda$ dominates a configuration $\mu$ if $n(\mu)$ can be derived from $n(\lambda)$ by applying a sequence of squeezing operations. When FQH wavefunctions,  equivalent to Jack polynomials are expanded in the occupation-number basis the only configurations with non zero weight are the dominant configuration and those derived from this via squeezing operations. This is also true of FQH states which are equivalent to the product of Jack polynomials and some power of Vandermonde determinants. Recent work  \cite{ar11,be11} has focused on the form of squeezing operations required for dealing with spinful states.

As a consequence of the pairing structure that we described in section \ref{sec:pairing_structure}  we will demonstrate that the difference between the HR  and JSS ground states can be described by an excitation of two neutral fermions of opposite spin at total momentum $k = 0$  in the corresponding root configurations. We can start from the neutral excitation spectrum of the JSS state in the thermodynamic limit with quasiparticle-quasihole minimum as sketched on the left of Fig.1. The spectrum is completely gapped from the ground state, Eq.(\ref{first}), with root configuration on a sphere given by
$(\bar{2}00\downarrow 0 \uparrow 0 \downarrow \ldots 0 \uparrow 00 \bar{2})$.
By $\bar{2}$ we denote a spin-singlet pair on a single orbital. The flux/
particle number $(N_{\phi}/N)$ relationship is $N_{\phi} = 2 N - 4$. We expect that by changing (decreasing) the $V_0$ component of the pseudo-potential series $\{V_0, V_1,0,0,\ldots\}$, $V_0, V_1 > 0$ the system will become gapless and described at $V_0 = 0$ by the HR state with excitation spectrum sketched on the right of Fig. 1 with root configuration $(\bar{2}000\bar{2}000\ldots\bar{2}000\bar{2})$
with the same flux/number of particles relationship. As we know from the previous analysis \cite{rg} the branch of gapless excitations of the HR state is described by neutral fermions (excitations due to unpaired particles in the BCS state). The neutral fermions exist \cite{rg} in the JSS state and it is this gapped branch around $k = 0$ that becomes gapless at the critical point. It is thus to be expected that at the transition the pair of neutral fermions of opposite spin, each of momentum $k = 0$ become part of the ground state configuration and description. Indeed we can convince ourselves by looking at the root configurations of the JSS and HR states that they differ by the excitation of two neutral fermions with opposite spin. Each bulk
spin singlet pair in the HR state becomes set apart by one orbital in the root
configuration of the JSS state. Opposite spin thus carry opposite momentum, but due to the requirement of inversion symmetry wrt the equator and the constraint on the flux/number of particle ratio (charge neutrality) the boundary configurations do not change and the difference between the two states may appear to us as some kind of boundary excitations in a uniform state (the JSS state). But as we already explained essentially the difference between the HR and JSS phase can be described by the state of two neutral fermion bulk excitations in their respective ground states.

\begin{figure}
\centering
\includegraphics[scale=0.35]{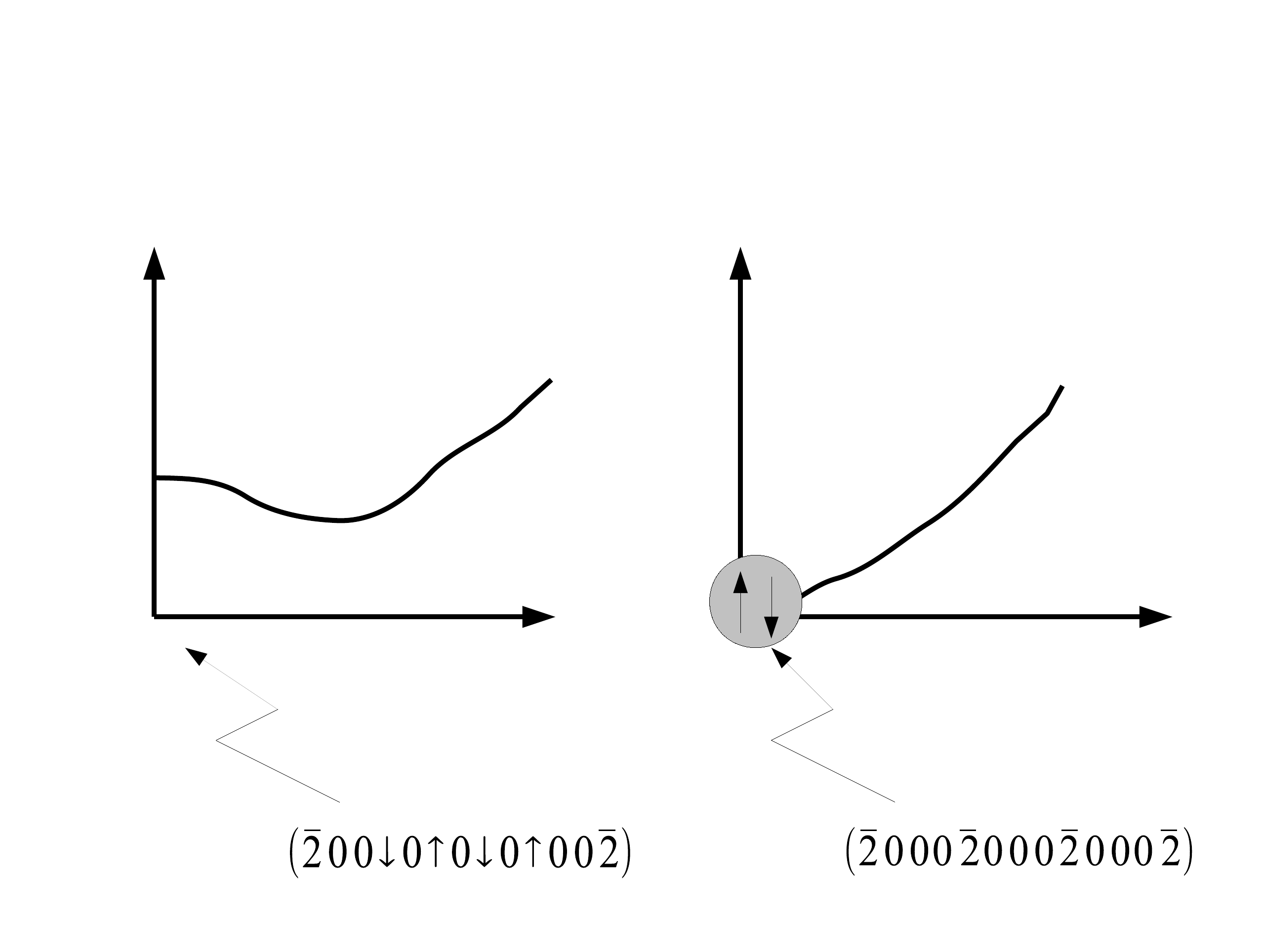}
\caption{The sketches of excitation spectra of Jain spin singlet (left panel) and Haldane Rezayi state (right panel) with
respective root configurations of the ground states.}
\end{figure}

\begin{figure}
\centering
\subfloat{
 \includegraphics[scale=0.5]{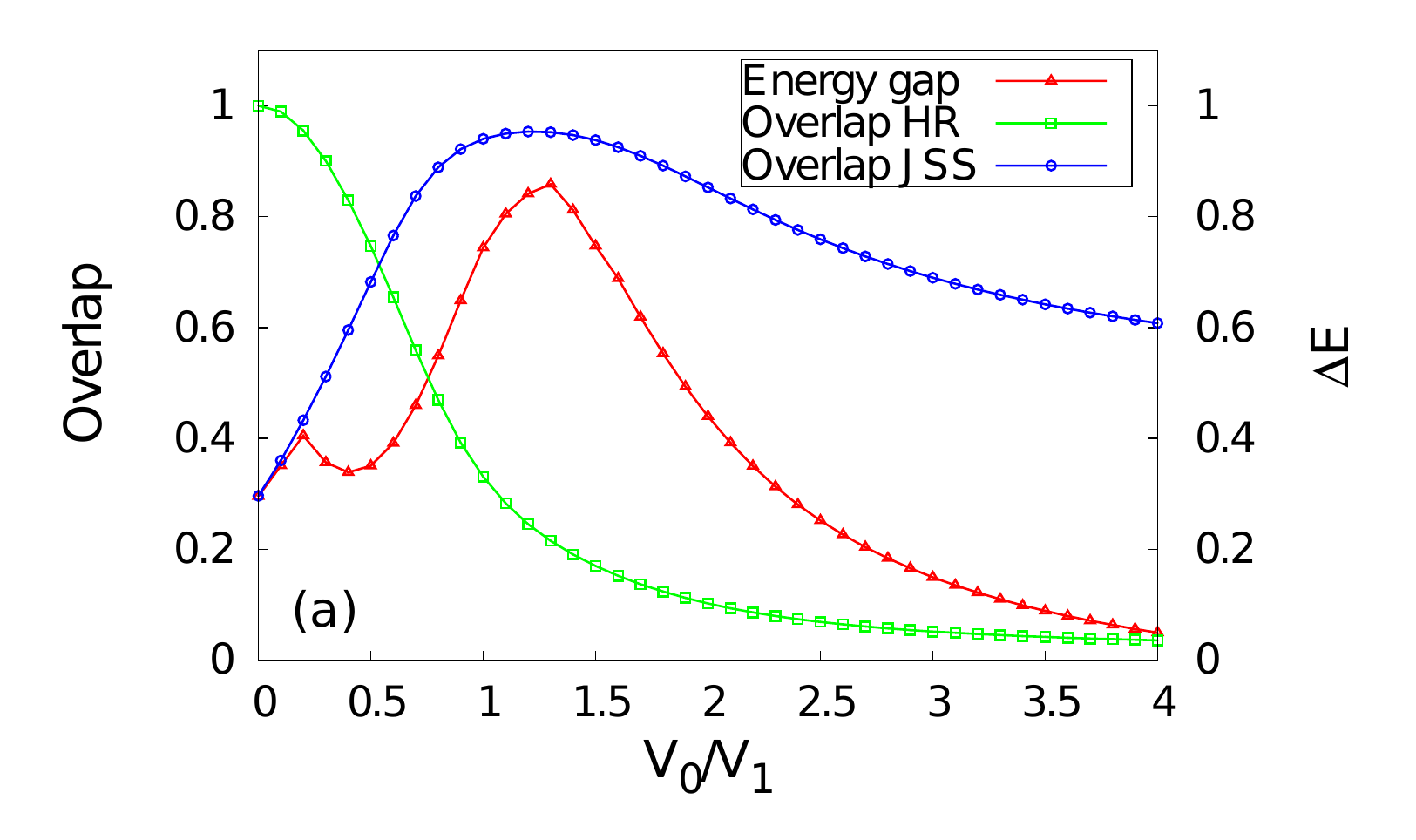}
}\\
\subfloat{
  \includegraphics[scale=0.5]{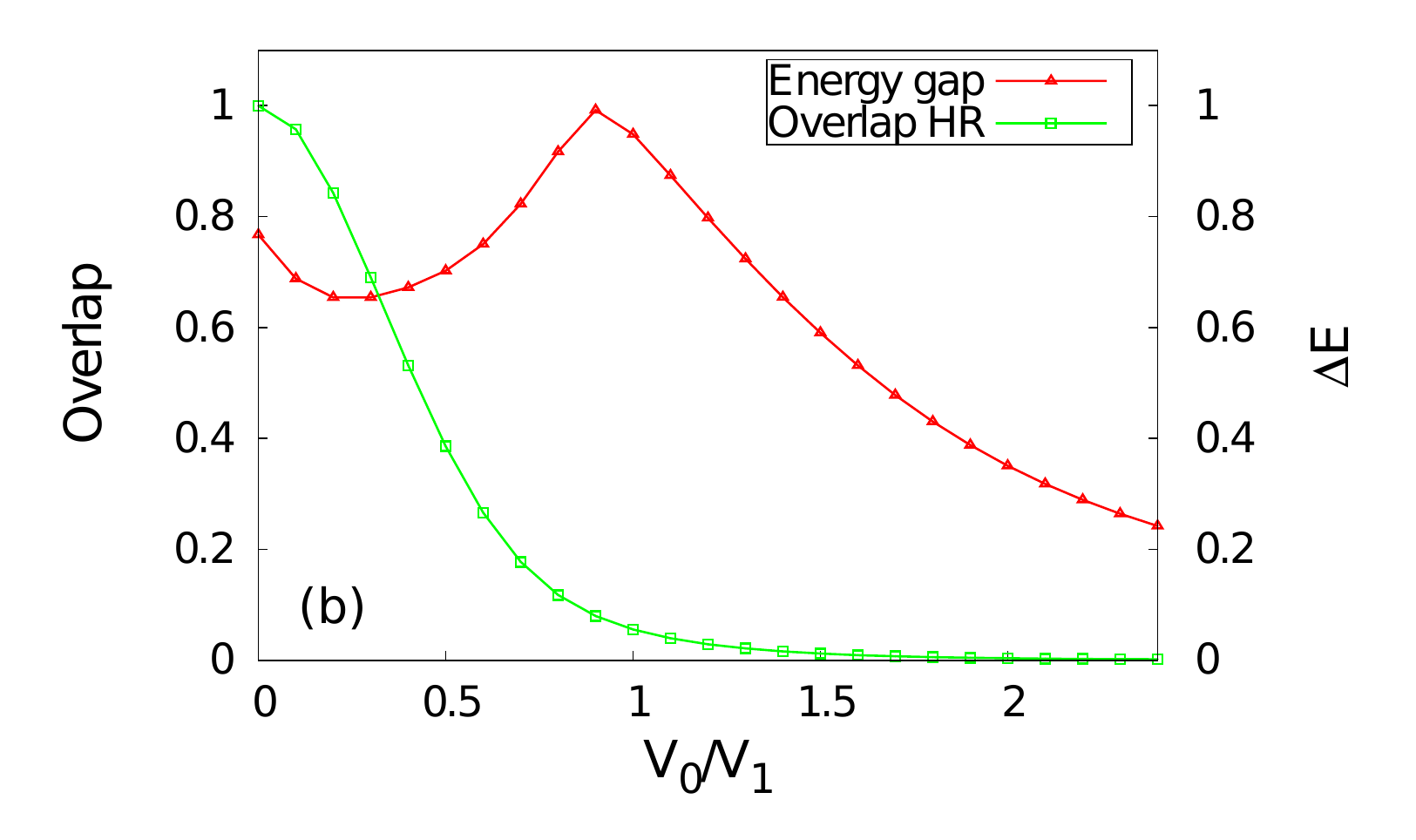}
}
\caption{Excitation gap and overlaps of the ground state for a two-body interaction Hamiltonian for different $\frac{V_0}{V_1}$ for (a) $N=10$ and (b) $N=12$ (see \cite{bwj} for plots of N=8 case). No overlap data are available for the JSS for $N = 12$.}\label{fig:gap_overlap}
\end{figure}

\section{Numerical calculations}\label{sec:num_calcs}
To verify the state on the weak paring side of the transition (HR state) is indeed a JSS state we obtain the ground state of the relevant interaction Hamiltonian and compare this to the explicitly constructed JSS state. The two body interaction here consists of a hollow-core interaction ($V_1 = 1$) along with a varying strength hard-core interaction ($V_0 > 0$).
Constructing the JSS wavefunction is very computationally intensive and $N=10$ was the largest we could construct. This is somewhat smaller than what has been achieved for spin polarized systems. This is because changing the order of application of the projection operator no longer results in a good approximation as it does in the spin polarized case (appendix \ref{ap1}). Figure \ref{fig:gap_overlap} shows the results of these calculations for the $N=10$ and $N=12$ cases. As expected, as $V_0$ is increased the overlap with the HR state decays. For $N=10$ where we could construct the JSS state we see that as $V_0$ is increased the overlap with this state increases to almost unity before starting to decay. This is a strong indication that this is indeed the JSS state on the week pairing side of the transition. In both cases the energy gap also shows a peak near where we expect the JSS state to be which is consistent with this picture.

\section{Spin polarized case}\label{sec:spin_polarised_case}
 In the following we will discuss spin polarized Jain states and their relationship to the non-unitary states. This subject is well studied, especially the case of bosons at filling factor 2/3 and related non-unitary state so-called Gaffnian state\cite{Yoshioka88,Simon07b}, and our focus here will be underlying pairing structures in these states. The root configurations of these two states, Jain and Gaffnian, are well known \cite{nr} and their pairing structure can be probed. We will see that also in this case, as the difference of two states, two neutral excitations exist that are spread out over the whole system. Due to the equivalence of North and South poles on the sphere (with magnetic monopole in its center) i.e. symmetry under inversion on the finite interval of angular momentum states of any quantum Hall state and as a consequence of the bulk neutral excitations ``edge decorations" - special decorations on the ends (North and South pole) appear as we find in the case for the JSS state.
 This neutral rearrangement and edge decorations can be seen in the root configuration of the Jain state: (2010110102) wrt that of the Gaffnian state
(2002002002) (for the sake of simplicity we displayed the root configurations for
only eight particles).
 This can also be seen in Jain states that need more than two LLs for their construction. Each new LL contributes a new pair of neutral excitations with respect to non-unitary partner states \cite{nr}. To understand the origin of this behavior that may stem from a pairing structure in Jain states, we begin with the definition of the Gaffnian wavefunction of bosons at 2/3:
\begin{equation}
\Psi_{Gf} = {\cal S}( \Psi_{221} \mathrm{perm}\left(\frac{1}{z_{\uparrow} - z_{\downarrow}}\right))
\label{Gaffnian}
\end{equation}
In constructing this state we first divide the electrons into two groups of up $(\uparrow)$ and down
$(\downarrow)$ pseudospin. In the definition Eq.(\ref{Gaffnian}), $\mathrm{perm}$ denotes the permanent which is for a $M\times M$ $\cal{M}$ matrix ${\rm perm}\left(\cal{M}\right)=\sum_{p\in \mathcal{S}_M}\prod_{k=1}^M\,\cal{M}_{k,p(k)}$.
 $\Psi_{221}$ is the well known notation of Halperin states for which we have
\begin{equation}
\Psi_{221} = \prod_{i<j} (z_{i \uparrow} - z_{j \uparrow})^2 \prod_{l<m} (z_{l \downarrow} - z_{m \downarrow})^2 \prod_{p<q} (z_{p \uparrow} - z_{q \downarrow}). \label{Halperin}
\end{equation}
In the following we will use
\begin{equation}
(z_{\sigma} - z_{\sigma'})^m,
\end{equation}
where $m$ can be a fraction and $(\sigma, \sigma') = (\uparrow,\uparrow),(\downarrow, \downarrow),$ or $(\uparrow, \downarrow)$ as a shorthand notation for any of the three factors in Eq.(\ref{Halperin}). The overall symmetrization operator, ${\cal S}$ in Eq.(\ref{Gaffnian}), is necessary to produce a state of polarized bosons.

To display the pairing structure related to the previous discussion of the HR state we will separate out the charge part, i.e. the part blind to pseudospin:
\begin{eqnarray}
\Psi_{Gf} =
 {\cal S} (\prod(z - z)^{3/2} \times \nonumber \\(z_{\uparrow} - z_{\uparrow})^{1/2}(z_{\downarrow} - z_{\downarrow})^{1/2} \frac{1}{(z_{\uparrow} - z_{\downarrow})^{1/2}} \mathrm{perm}\left(\frac{1}{(z_{\uparrow} - z_{\downarrow})}\right)) \nonumber \\ \label{separate1}
\end{eqnarray}
where $\prod (z - z)^{3/2}$ denotes the product of all possible pairs:
\begin{equation}
\prod (z - z)^{3/2} = (z_{\uparrow} - z_{\uparrow})^{3/2}(z_{\downarrow} - z_{\downarrow})^{3/2}(z_{\uparrow} - z_{\downarrow})^{3/2}.
\end{equation}
Due to the equality given in \cite{hr},
\begin{equation}
\Psi_{11-1} \mathrm{perm} \left(\frac{1}{z_{\uparrow} - z_{\downarrow}}\right) = \det\left(\frac{1}{(z_{\uparrow} - z_{\downarrow})^2}\right),
\end{equation}
we can rewrite the Gaffnian as
\begin{eqnarray}
\Psi_{Gf} =
 {\cal S} (\prod(z - z)^{3/2} \times \nonumber \\\frac{(z_{\uparrow} - z_{\downarrow})^{1/2}}{(z_{\uparrow} - z_{\uparrow})^{1/2}(z_{\downarrow} - z_{\downarrow})^{1/2}}  \det\left(\frac{1}{(z_{\uparrow} - z_{\downarrow})^2}\right)) . \label{separate2}
\end{eqnarray}
Thus  a possible interpretation of the Gaffnian state is that it represents a HR pairing state of neutral semions, semions because we have taken in front the factor $\prod (z - z)^{3/2}$ that describes the charge part. The original semions that paired in by way of a permanent in the usual definition (Eq.(\ref{separate1})) have relative fermionic statistics with respect to the new semions of Eq.(\ref{separate2}).

We can try to extend our pairing arguments from spin singlet HR and Jain state to Gaffnian and Jain state at 2/3 (at 2/5 in the case of fermions). We expect that the Jain  state at 2/3 can be viewed as an underlying state of weakly paired semions as in the following expression (we  neglect the projection to the LLL in the following)
\begin{eqnarray}
\Psi_{Jain} =
 {\cal S} (\prod(z - z)^{3/2} \times \nonumber \\\frac{(z_{\uparrow} - z_{\downarrow})^{1/2}}{(z_{\uparrow} - z_{\uparrow})^{1/2}(z_{\downarrow} - z_{\downarrow})^{1/2}}  \det\left(\frac{z^*_{\uparrow} - z^*_{\downarrow}}{z_{\uparrow} - z_{\downarrow}}\right)). \label{separateJ}
\end{eqnarray}
Due to the previously proven identity (Eq.(\ref{identity})),
\begin{equation}
\chi_2 = \frac{\chi_1}{\chi_{110}} \det\left(\frac{z^*_{\uparrow} - z^*_{\downarrow}}{z_{\uparrow} - z_{\downarrow}}\right), \label{id2}
\end{equation}
we can rewrite Eq.(\ref{separateJ}) as
\begin{eqnarray}
\Psi_{Jain} =
 {\cal S} (\prod(z - z)^{3/2} \times \nonumber \\\frac{(z_{\uparrow} - z_{\downarrow})^{1/2}}{(z_{\uparrow} - z_{\uparrow})^{1/2}(z_{\downarrow} - z_{\downarrow})^{1/2}}  \frac{\chi_2 \chi_{110}}{\chi_1})\nonumber \\
 = {\cal S}(\chi_1 \chi_2) = \chi_1 A(\chi_2) = \chi_1 \chi_2, \label{separateJJ}
\end{eqnarray}
as we anticipated. The last identity in which $A$ is antisymmetrizer follows from the antisymmetry already encoded in $\chi_2$ under exchange of any $i$ and $j$. 
Moreover we can start from the definition of the bosonic Jain state,
\begin{equation}
\Psi_{Jain} = \chi_2 \chi_1,
\end{equation}
use the same identity in Eq. (\ref{id2}), and conclude that
\begin{eqnarray}
\Psi_{Jain} &=&  \det\left(\frac{z^*_{\uparrow} - z^*_{\downarrow}}{z_{\uparrow} - z_{\downarrow}}\right) \times \nonumber \\
&& (z_{\uparrow} - z_{\uparrow}) (z_{\downarrow} - z_{\downarrow})  (z_{\uparrow} - z_{\downarrow})^2 \label{jseparate},
\end{eqnarray}
i.e. come to an expression for $\Psi_{Jain}$ in terms of two groups of particles. As we emphasized below Eq.(\ref{identity}), the division between ups and downs in Eq.(\ref{jseparate}) is arbitrary and we do not have a regular paired state with a charge part clearly separated form a pairing function. As before, but without the need for symmetrizer ${\cal S}$ we have
\begin{eqnarray}
\Psi_{Jain} =
 \prod(z - z)^{3/2} \times \nonumber \\\frac{(z_{\uparrow} - z_{\downarrow})^{1/2}}{(z_{\uparrow} - z_{\uparrow})^{1/2}(z_{\downarrow} - z_{\downarrow})^{1/2}}  \det\left(\frac{z^*_{\uparrow} - z^*_{\downarrow}}{z_{\uparrow} - z_{\downarrow}}\right).
\end{eqnarray}

Therefore we conclude that the Jain state at 2/3 can be (to a certain degree) viewed as a topological superconductor of anyons in a weak pairing phase. It  is not obvious what the physical consequences of such a statement are. The pairing is  very much disguised.  We may also talk about neutral fermions
and their pairing, but there is no simple relationship between them and the underlying particles - in this case bosons.

Edge decorations in the root configuration of the Jain state in comparison with the Gaffnian clearly point to the presence of neutral excitations that follow from pair breaking.
To understand better how edge decorations are connected with the pairing structure in Gaffnian and Jain states that we demonstrated previously in Eqs.(\ref{separate1},\ref{separate2},\ref{separateJ},\ref{separateJJ}) we will take out ${\cal S}$ (symmetrizer) in the definition of the Gaffnian (Eq.(\ref{Gaffnian})). As a result we get a spinful state with root configuration: $(\bar{2}00\bar{2}00\bar{2}00\bar{2})$ where $\bar{2}$ represents a spin singlet on a single orbital. (This is analogous to the HR case.) We may imagine pair-breaking neutral excitations with spin which would lead to root configurations  of the following form $(\bar{2} 0 \uparrow 0 \downarrow  \uparrow 0 \downarrow 0 \bar{2})$, but this would be too restrictive to describe the root configuration of a Jain state which is ferromagnetically ordered with the total projection along the quantization axis equal to zero in the pseudospin space : $(2010110102)$. We can convince ourselves of this particular ferromagnetic ordering by analyzing the expression Eq.(\ref{separateJ}) for the Jain state. Nevertheless we see the similarity between pair-breaking neutral excitations that carry spin and quasiparticle-qusihole excitations \cite{bhqe} on both ends of the Jain state. Here quasiparticle-quasihole excitations correspond  to neutral fermions in the HR and JSS case. Instead of a pair of neutral fermions of opposite spin in the polarized Jain case we have a quadrupolar \cite{zh} excitation, two quasiparticle-quasihole pairs that are spread out over the ground state. Namely we need two (neutral) dipoles of
corresponding but opposite momenta to make $k = 0$ excitation that falls down in energy when we are approaching the critical Gaffnian state. The situation is similar to the spin-singlet HR and JSS case with opposite spin neutral fermions as sketched in Fig. 1.

With all said about Jain states we can expect that IQHE wavefunctions that describe non-interacting fermions can be described as some kind of weakly paired topological superconductors where the extremely
weak pairing of time reversal symmetry breaking $d$-wave, which is just a phase, goes into the description
of fermionic correlations between different LLs.
As we already demonstrated  the Slater determinant of two filled LLs can be written as
\begin{equation}
\chi_2 = \det\left(\frac{z^*_{\uparrow} - z^*_{\downarrow}}{z_{\uparrow} - z_{\downarrow}}\right) (z_{\uparrow} - z_{\downarrow}). \label{2ll}
\end{equation}
We emphasize that the division between ups and downs is arbitrary; the only requirement is the equal number of ups and downs i.e. the total even number of fermionic particles. The  factor, $ (z_{\uparrow} - z_{\downarrow})$, similar to the Jastrow-Laughlin factor but not the same, carries the information of the filling factor i.e. from the number of flux quanta that particles experience $N_{\phi}^{\uparrow} = N_{\phi}^{\downarrow} = N/2 - 2$ we can read off the filling factor, $\nu = 2$.

The interesting question is what is the relationship between weakly paired $d$-wave superconductors and the topological insulator i.e. IQHE with Chern number equal to 2. This question is highly relevant in the context of fractional Chern insulators\cite{Haldane88} (i.e. FQHE without magnetic field) with Chern number larger than 1 \cite{qi2011,wang2011,wang2012}. Besides a relationship between bulk Hamiltonians defined on a lattice as demonstrated in Ref.{\cite{lau}, there is obvious similarity in the edge theories, both are made up of
two Dirac fermions \cite{rg}, which expressed in Majoranas, represent a theory with $SO(4)$ symmetry which is equivalent to $SU(2) \times SU(2)$ symmetry. We may ask what is the symmetry of bulk $d$-wave Hamiltonians in order to identify the degrees of freedom which are transformed under the symmetry. First there is obvious spin rotation symmetry, $SU(2)_{spin}$, due to the underlying spin degree of freedom in the Hamiltonian; the ground state wavefunction,
\begin{equation}
\Psi_d = \det\left(\frac{z^*_{\uparrow} - z^*_{\downarrow}}{z_{\uparrow} - z_{\downarrow}}\right),
\end{equation}
is a spin-singlet - eigenstate of $SU(2)_{spin}$ by being a collection of BCS spin-singlet pairs. Second, besides particle-hole symmetry, there is no additional internal symmetry in the BCS Hamiltonian. Only in its ground state wavefunction is the number of complex-conjugated variables and the number of non-conjugated (``LL index") expected to be the same or expressed in an equivalent way their difference should be conserved. Hence we may talk about an internal $U(1)$ symmetry.
What we can conclude is that the symmetry that is present in the bulk is enlarged at the edge to $SU(2)_{spin} \times SU(2)_{internal}$.

On the other hand in the case of IQHE at $\nu = 2$ at the edge we may talk certainly about a symmetry that acts on the LL index in parallel with the spin symmetry on the edge of $d$-wave superconductors. Therefore on the edge we have a $SU(2)_{LL\;index} \times SU(2)_{internal}$ symmetry. (Note that here $SU(2)_{internal}$ should not be identified  with the one in the context of the $d$-wave superconductor.) There are no explicit degrees of freedom in the bulk that would correspond to or lead to $SU(2)_{internal}$ symmetry on  the edge. Interestingly  the bulk ground state wavefunction has the form which can be seen in Eq.(\ref{2ll}) that it is invariant under arbitrary assignment of ups and downs. Eq.(\ref{2ll}) relates the ground state wavefunction of $d$-wave superconductors and IQHE at $\nu = 2$ and therefore indicates a pairing structure in IQHE wavefunctions. There is no pseudospin degree of freedom in IQHE  (Hamiltonian) in the bulk, but the ground state wavefunction looks as if there is an additional ferromagnetically ordered pseudospin degree of freedom next to the LL index. And the symmetries related to these structures exist on the edge.

 Therefore IQHE and polarized FQHE states underlie pairing construction which incorporates the ``right" mutual statistics of constituents that is achieved by their $d$-wave pairing. At the same time their construction incorporates an explicit projection to ferromagnetic i.e. one-component state so that the paired nature is suppressed. In this way latent, pseudospin degrees of freedom that are paired in the ground state wave functions  appear in the root configurations of the model wavefunctions and on the edge in the way of
enlarged symmetry.

\section{Conclusions}\label{sec:conclusions}

Haldane-Rezayi and Jain spin singlet states are canonical examples of $d$-wave pairing of FQHE wave functions. We explicitly showed $d$-wave pairing in the case of the JSS state. The root configuration of the JSS state was derived in which we could recognize the role of neutral fermion pairs in the transition from the JSS to the HR state. We demonstrated this transition in an exact diagonalization study. Besides its intrinsic interest the study enabled us to make parallels and conclusions concerning polarized FQHE and IQHE states. We found the presence of the $d$-wave pairing in
these states although it is suppressed due to their one-component nature.

\begin{acknowledgments}
We thank V. Gurarie for useful discussions. A.S. thanks Princeton University for generous hosting. A.~S. was supported by Keck grant and N.~R. was supported by  NSF CAREER DMR-095242, ONR - N00014-11-1-0635, Packard Foundation and Keck grant.  I.~V. and M.V.M. were supported by the Serbian Ministry of Education and Science under project No. ON171017. The authors also acknowledge support form the bilateral MES-CNRS 2011/12 program. This research was supported in part by the National Science
Foundation under Grant No. NSF PHY11-25915; M.V.M. acknowledges the hospitality of KITP, Santa Barbara.
\end{acknowledgments}

\appendix
\section{Numerical construction of composite fermion wavefunctions}\label{ap1}
First we discuss the construction of lowest Landau level spin polarized composite fermion wavefunctions of the form
$$\phi = \mathcal{P}_{LLL} \left[ \chi_1^{2p} \chi_n \right] $$
In \cite{jk} it was demonstrated that when constructing wavefunctions of this form the Jastrow factor $\chi_1^{2p}$ can be moved inside the determinant coming from $\chi_n$. The LLL projection can then be performed before taking the determinant. In addition, analytical expressions for the application for the LLL projection operator can be derived. In this manner the computational cost of constructing such wavefunctions is dramatically reduced.

Extending this, it was discovered that this method can be applied even for cases where the wavefunction in question does not have this form. For example, the bosonic wavefunctions considered in Ref.\cite{jr} that are associated to CF state at filling factor $\nu = \frac{n}{n+1}$ fall into this category.
$$\phi_B = \mathcal{P}_{LLL} \left[ \chi_1 \chi_n \right] $$
It was shown that this wavefunction can be approximated well with
$$\phi_B^{\prime} = \chi_1^{-1} \mathcal{P}_{LLL} \left[ \chi_1^2 \chi_n \right] $$
which is amenable to the technique from \cite{jk}. The overlap for $N=8$ is $\abs{\braket{\phi_B}{\phi_B^{\prime}}}^2=0.9820$ \cite{jr}.

In the case of the JSS wavefunction it was hoped that a similar method could be applied. We constructed the wavefunctions
$$\phi_{JSS}^{\prime} = \chi_{001}^{-1} \mathcal{P}_{LLL} \left[ \chi_1^2 \chi_2 \right] $$
and
$$\phi_{JSS}^{\prime \prime} = \chi_{110} \mathcal{P}_{LLL} \left[ \chi_1 \chi_2 \right] $$
However it was found that these do not offer good approximations of the JSS state even for small systems. The overlaps with the JSS state for $N=8$ are $\abs{\braket{\phi_{JSS} }{\phi_{JSS}^{\prime}}}^2 = 0.790$ and $\abs{\braket{\phi_{JSS} }{\phi_{JSS}^{\prime\prime}}}^2 = 0.792$. Note that for $\phi_{JSS}^{\prime \prime}$ the term inside the projection is not of the form $ \chi_1^{2p} \chi_2 $ and thus is not amenable to the technique described in \cite{jk}. However this wavefunction is still less computationally intensive to construct than $\phi_{JSS}$ (Eq. \ref{first}) since the application of the projection operator before performing the product operation makes this operation much less demanding.

\end{document}